\documentclass[aps,twocolumn,pra,tightenlines,floatfix,nofootinbib]{revtex4}
\usepackage[dvips]{graphicx}
\usepackage[english]{babel}
\usepackage{amsmath}
\usepackage{amssymb}
\usepackage{bm}
\usepackage{color}
\usepackage{subfigure}

\newcommand{\bx}{\mathbf{x}}
\newcommand{\bk}{\mathbf{k}}
\newcommand{\bq}{\mathbf{q}}
\newcommand{\nn}{\nonumber}
\newcommand{\om}{\omega_n}
\newcommand{\dg}{\dagger}
\newcommand{\dm}{\delta\mu}
\newcommand{\ua}{\uparrow}
\newcommand{\da}{\downarrow}
\newcommand{\mf}{{sc}}

\newcommand{\pg}{{pg}}
\newcommand{\be}{\begin{eqnarray}}
\newcommand{\ee}{\end{eqnarray}}
\newcommand{\ea}{\end{array}}
\newcommand{\ba}{\begin{array}}

\begin{document}
\title{BCS-BEC crossover of Spin Polarized Fermi Gases with Rashba Spin-Orbit Coupling}

\author{Chang-Yan Wang}
\affiliation{College of Physical Science and Technology,
Sichuan University, Chengdu, Sichuan 610064, China}
\author{Yan He}
\affiliation{College of Physical Science and Technology,
Sichuan University, Chengdu, Sichuan 610064, China}
\date{\today}

\begin{abstract}
We study the BCS-Bose Einstein Condensation (BEC) crossover of a three dimensional spin polarized Fermi gas with Rashba spin-orbital-coupling (SOC). At finite temperature, the effects of non-condensed pairs due to the thermal excitation are considered based on the $G_0G$ pair fluctuation theory. These fluctuations generate a pseudogap even persistent above $T_c$. Within this framework, the Sarma state or the spin polarized superfluid state and polarized pseudogap state are explored in detail. The resulting $T_c$ curves show that the enhancement of pairing due to the SOC roughly cancels out the suppression of pairing due to the population imbalance. Thus we observed that in a large portion of the parameter space, the polarized superfluid state are stabilized by the SOC.
\end{abstract}
\maketitle

\maketitle

\section{Introduction}

Ultracold Fermi gases with tunable interactions has been the focus of a lot of experimental and theoretical works.
The advantage of cold Fermi gases is that they are much more controllable than other ordinary condensed matter materials. The interaction between the fermions usually has very short effective range and can be characterized by one parameter, the s-wave scattering length $a$, or more conveniently, dimensionless scattering length $1/(k_Fa)$, with the Fermi momentum $k_F$. By applying a external magnetic field, one can tune the Zeeman splitting energy between the bound state of closed channel and the continuum threshold of the open channel, which will lead to the so-called Fashbach resonance\cite{Jin4,Grimm2,KetterleV} when the above two energy levels line up. This phenomenon allows one to tune $1/(k_Fa)$ from very negative to a large positive number. Correspondingly, the interatomic interaction varies from a weak attraction to a very strong attraction. Therefore, the cold Fermi gases provide an experimental realization of the early theoretical ideas of BCS-BEC crossover\cite{Eagles,Leggett,NSR,MR93} and are also very useful for testing the many-body theories.

A natural interesting point in the crossover is the so-called unitary limit when the scattering length diverges. The unitary limit of Fermi gases is intrinsically strong correlated and there is no small parameter to expand with. The BCS-BEC crossover theory describes the loosely bound Cooper pairs evolve into tightly bound bosonic pairs with increasing attractive interaction strength. It captures the two weakly interacting limits and can also give quantitative account of the unitary limit in between. When considering the finite temperature physics of the crossover, the contribution of the non-condensed pairs become important due to the stronger than BCS attraction and these effects must be considered in a way consistent with the BCS ground state.

Soon after the experimental evidence of the fermionic superfluidity was achieved, one was able to adjust the number density of spin up and spin down particles. Since the singlet pairing requies equal number of spin up and down particles, this spin polarized Fermi gases generated a lot of possible novel phases\cite{stringari-RMP}, such as Sarma or polarized superfluid phase\cite{Sarma,V-Liu}, polarized pseudogap phase, Larkin-Ovchinnikov-Fulde-Ferrell (LOFF) phase\cite{FFLO,LOFFlong,LOFF-RMP}, or the phase separations between different phases\cite{ChienPRL,SR06}. The stability of these phases and the transitions between them have also been studied by many authors\cite{Stability,Pao}.

In recent years, another important breakthrough in experiments is the realization of the synthetic non-Abelian gauge field and the spin-orbital-coupling (SOC) in cold atomic gases\cite{spielman}. By applying two counter propagating Raman laser beams and a transverse Zeeman field to Boson atoms with multiple components, a synthetic SU(2) non-Abelian gauge field can be generated by appropriate choice of the laser frequency and Zeeman splitting. One can show that the low energy effective theory contains a Raman type of SOC terms. Similar scheme can also be applied to the fermion atoms, and the SOC of Fermi gases has already been generated in experiments\cite{SOC-SX,SOC-MIT}. The SOC of Fermi gases opens up the possibilities to explore many novel physics in cold atoms. It also provides us yet another way to achieve the BCS-BEC crossover. As pointed out in \cite{Shenoy}, with Rashba type of SOC, there appears a new type of two-body bound state of fermion pairs even for $a<0$, while two-body bound state does not exit in the BCS side without SOC. This suggests that the pairing strength between fermion pairs are enhanced by the Rashba SOC term. Therefore, increasing the SOC coupling is equivalent to push the Fermi gases to its deep BEC limit, thus is another way to get BCS-BEC crossover.

There already appears a lot of theoretical works on the SOC Fermi gases with or without population imbalance\cite{zhai-review}. However, other than a few exceptions\cite{hu}, most of these works are based on the mean field theory which should be more appropriate for the deep BCS or very low temperature cases. The reason is that, as we mentioned before, in the unitary limit or BEC side of the crossover, the attraction is strong enough to support two-body bound pairs. At the finite temperature, there will be substantial amount of non-condensed pairs which generate an energy gap in single particle spectrum. In contrast to the superfluid order parameter, this pseudogap does not signal symmetry breaking. Thus it can be persistent above $T_c$ which leads to a non-Fermi liquid normal state. Different pair fluctuation theories differentiate themselves from each other in the detailed form of pair propagator or T-matrix\cite{Haussmann,Milstein,Randeria97}. In this paper, we follow the so-called $G_0G$ pair fluctuation theory\cite{Ourreview}, which is partly inspired by the early work of Kadanoff and Martin\cite{KM}. The advantage of this theory is its consistency with the BCS ground state and its numerical calculability. With both SOC and spin polarization, there maybe emerges a lot of exotic phases. To map out the whole phase diagram will be a enormous task. In this paper, we only consider the simplest Sarma state or polarized superfluid state and polarized pseudogap state. To avoid other possible phases, we will mostly confine ourself in the parameter space where the polarized superfluid state is stable.

This paper is organized as follows. In section \ref{mf}, we present the mean field theory of population imbalanced
Fermi gases with Rashba SOC. Then we generalized the mean field theory to include the $G_0G$ pair fluctuation effects in section \ref{pair}. In section \ref{numerics}, we present the numerical results and discussion bases on the previous theoretical formalism. We conclude in section \ref{conclu}.

\section{Mean field theory}
\label{mf}

In this section, we consider the BCS mean field theory of population imbalanced Fermi gases with Rashba type of SOC. We will see that the gap equation can be rewritten in the T-matrix form, according to the Thouless criterion. Through this form, it is easy to generalize the BCS mean field theory to include the $G_0G$ pairing fluctuation effects.

The system of two-component population imbalanced Fermi gases with Rashba SOC across a Feshbash resonance can be described by the Hamiltonian:
\be
H&=&\int d^3\bx\psi^\dg(x)[-\frac{\nabla^2}{2m}-\mu-\dm\sigma_z+\mathcal{H}_{so}]\psi(x)\nn\\
&&+g\int d^3\bx\psi_\ua^\dg(\bx)\psi_\da^\dg(\bx)\psi_\da(\bx)\psi_\ua(\bx),\label{Ham}
\ee
where $\psi^\dg(x)=(\psi_\ua^\dg(x),\psi_\da^\dg(x))$ is the creation operator, $\mu=(\mu_\ua+\mu_\da)/2$ is the average of the chemical potentials, $\delta\mu=(\mu_\ua-\mu_\da)/2$ is the chemical potential difference, $\mathcal{H}_{so}=-i\lambda(\sigma_x\partial_x+\sigma_y\partial_y)$ is the Rashba SOC term, $\sigma_x, \sigma_y, \sigma_z$ are the three Pauli matrices, $g<0$ is the bare interaction strength of s-wave attractive interaction.
In this paper, we take $\hbar=k_B=1$ for convenience.

It is more convenient to introduce the Nambu spinor $\Psi^\dg=(\psi_\ua^\dg, \psi_\da^\dg, \psi_\ua, \psi_\da)$. Because the SOC interaction term explicitly depends on the spin indices, the Nambu spinor has 4 components, mean field Hamiltonian and Green's functions in Nambu space are 4 by 4 matrices. Then the imaginary-time Green's function can be written as:
\be
\mathcal{G}(\tau,\bx)&=&-\langle T_\tau\Psi(\tau,\bx)\Psi^\dg(0,0)\rangle\nn\\
&=&\left[\ba{cc}G(\tau,\bx)&F(\tau,\bx)\\\tilde{F}(\tau,\bx)&\tilde{G}(\tau,\bx)\ea\right],
\ee
where $T_\tau$ is the time order operator. After Fourier transformation, we find the Green's function in frequency-momentum space:
\be
\mathcal{G}(K)=
\left[\ba{cc}G(K)&F(K)
\\\tilde{F}(K)&\tilde{G}(K)\ea\right],
\ee
Here $K=(i\om,\bk)$, $\om=(2n+1)\pi T$ is the Matsubara frequency for fermion.
And its matrix elements also satisfy the following relations
\be
\tilde{G}(K)=-G(-K)^T\label{Gt}\\
\tilde{F}(K)=-F(-K)^T\label{Ft}
\ee

Next we will derive the gap equation, number density equation, number density difference equation based on the mean field approximation $\Delta_{sc}=g\langle \psi_\da(\tau,\bx)\psi_\ua(\tau,\bx)\rangle$. In this paper, we only consider the pairing in the spin singlet channel. For convenience, we take $\Delta_\mf$ to be real, i.e. $\Delta_{sc}^*=\Delta_{sc}$.

By rewriting the Hamiltonian Eq.(\ref{Ham}) in the Nambu space and making the mean field approximation to the two-body interaction\cite{stoof}, we find that the inverse BCS propagator in frequency-momentum space can be expressed as
\be
\mathcal{G}^{-1}(K)=
\left[\ba{cc}G_0^{-1}(K)&-i\Delta_\mf\sigma_y\\
i\Delta_\mf\sigma_y&\tilde{G}_0^{-1}(K)\ea\right],
\ee
where
\be
G_0^{-1}(K)=i\om-(\xi_\bk-\dm\sigma_z)-\lambda(k_x\sigma_x+k_y\sigma_y),\nn\\
\tilde{G}_0^{-1}(K)=i\om+(\xi_\bk-\dm\sigma_z)-\lambda(k_x\sigma_x-k_y\sigma_y)\nn,
\ee
with $\xi_\bk=\bk^2/2m-\mu$.

A simple matrix inversion gives the following full BCS Green's function in Nambu space
\be
\mathcal{G}(K)=\left[\ba{cc}G(K)&F(K)
\\\tilde{F}(K)&\tilde{G}(K)\ea\right],
\ee
where $G(K)$ and $F(K)$ are 2 by 2 matrices with the following matrix elements
\be
G_{11}&=&\sum_{\gamma,\alpha}\Big\{\alpha\gamma[\xi_\bk\eta_\bk^2-\dm(\Delta_\mf^2+\xi_\bk^2)-\gamma\dm\xi_\bk E_\bk^\alpha]\nn\\
&&+\rho_\bk E_\bk^\alpha+\gamma\rho_\bk(\xi_\bk-\dm)\Big\}\frac{1}{4\rho_\bk E_\bk^\alpha(i\om-\gamma E_\bk^\alpha)},\nn\\
G_{12}&=&\sum_{\gamma,\alpha}\frac{\gamma\lambda(k_x-ik_y)(\alpha\xi_\bk^2+\alpha\gamma\xi_\bk E_\bk^\alpha+\rho_\bk)}{4\rho_\bk E_\bk^\alpha(i\om-\gamma E_\bk^\alpha)},\nn\\
G_{21}&=&\sum_{\gamma,\alpha}\frac{\gamma\lambda(k_x+ik_y)(\alpha\xi_\bk^2+\alpha\gamma\xi_\bk E_\bk^\alpha+\rho_\bk)}{4\rho_\bk E_\bk^\alpha(i\om-\gamma E_\bk^\alpha)},\nn\\
G_{22}&=&\sum_{\gamma,\alpha}\Big\{\alpha\gamma[\xi_\bk\eta_\bk^2+\dm(\Delta_\mf^2+\xi_\bk^2)+\gamma\dm\xi_\bk E_\bk^\alpha]\nn\\
&&+\rho_\bk E_\bk^\alpha+\gamma\rho_\bk(\xi_\bk+\dm)\Big\}\frac{1}{4\rho_\bk E_\bk^\alpha(i\om-\gamma E_\bk^\alpha)},\nn
\ee
\be
F_{11}&=&\sum_{\alpha,\gamma}-\frac{\alpha\gamma\lambda (k_x-ik_y)\Delta (\xi+\delta\mu)}{4\rho E_\bk^\alpha(i\om-\gamma E_\bk^\alpha)},\nn\\
F_{12}&=&\sum_{\alpha,\gamma}\frac{\gamma\Delta(\rho+\alpha\delta\mu^2)-\alpha\Delta\delta\mu E_\bk^\alpha}{4\rho E_\bk^\alpha(i\om-\gamma E_\bk^\alpha)},\nn\\
F_{21}&=&\sum_{\alpha,\gamma}-\frac{\gamma\Delta(\rho+\alpha\delta\mu^2)+\alpha\Delta\delta\mu E_\bk^\alpha}{4\rho E_\bk^\alpha(i\om-\gamma E_\bk^\alpha)},\nn\\
F_{22}&=&\sum_{\alpha,\gamma}\frac{\alpha\gamma\lambda (k_x-ik_y)\Delta (\xi-\delta\mu)}{4\rho E_\bk^\alpha(i\om-\gamma E_\bk^\alpha)}\nn
\ee
and the other two tilted Green's functions are given by Eq.(\ref{Gt}) and (\ref{Ft}).
Here for convenience, we introduce $\eta_\bk=\sqrt{\lambda^2(k_x^2+k_y^2)+\dm^2},\ \rho_\bk=\sqrt{\xi_\bk^2\eta_\bk^2+\Delta_\mf^2\dm^2}$. The qusi-particle energy is given by $E_\bk^\alpha=\sqrt{\xi_\bk^2+\Delta_\mf^2+\eta_\bk^2+2\alpha \rho_\bk}$ and $\alpha,\gamma=\pm$.

With the full BCS Green's function, we find the self-consistent gap equation by evaluating the expectation $\langle\psi_\da(\tau,\bx)\psi_\ua(\tau,\bx)\rangle$ as follows
\be\label{eq:gap}
\frac{\Delta_{sc}}{g}&=&-\frac{1}{2}\mathrm{Tr} \sum_{K}i\sigma_y F(K)\nn\\
&=&-\frac{1}{4}\sum_{\bk,\alpha}\frac{\partial E_\bk^\alpha}{\partial\Delta_\mf}[1-2f(E_\bk^\alpha)],
\ee
Here $\sum_K=T\sum_n\sum_{\bk}$ and $f(x)=1/[\exp(x/T)+1]$ is the Fermi-Dirac distribution function.

Similarly, the number density equation is
\be\label{eq:n}
n&=&n_\ua+n_\da=\mathrm{Tr}\sum_K G(K)\nn\\
&=&\frac{1}{2}\sum_{\bk,\alpha}\Big[1+\frac{\partial E_\bk^\alpha}{\partial\mu}[1-2f(E_\bk^\alpha)]\Big],
\ee
and number density difference equation is given by
\be\label{eq:dn}
\delta n&=&n_\ua-n_\da=\mathrm{Tr}\sum_K i\sigma_z G(K)\nn\\
&=&\frac{1}{2}\sum_{\bk,\alpha}\frac{\partial E_\bk^\alpha}{\partial\dm}[1-2f(E_\bk^\alpha)],
\ee
It should be noted that the above three equations can also be obtained directly by taking derivatives of the thermodynamic potential as in\cite{iskin}.

We have derived the gap equation from standard mean field theory calculations. The gap equation can also be obtained from a T-matrix point of view. The T-matrix can be thought as the proper pair propagator with external legs been amputated.  According to the Thouless criterion\cite{ThoulessCriterion}, the divergence of many-body T-matrix signals the instability of normal ground state, which leads to the condensation of fermion pairs and the formation of the superfluid ground state. The many-body T-matrix can usually be approximated by the summation of a series of ladder diagrams as
\be\label{eq:t-mat}
t(Q)=\frac{g}{1+g\chi(Q)},
\ee
where $\chi(Q)$ is the mean-field pair susceptibility\cite{Ourreview,hu}:
\be
\chi(Q)=\frac{1}{2}\mathrm{Tr}\sum_K  \tilde{G}_0(K-Q)i\sigma_y G(K)i\sigma_y
\ee
with $Q=(i\Omega_n,\bq)$, $\Omega_n=2n\pi T$ is the Matsubara frequency of bosons. We note that the pair susceptibility is made by one full and one bare Green's functions. As pointed by Kadanoff and Martin long time ago\cite{KM}, this is important to recover the BCS gap equation. The Thouless criterion require that the many-body T-matrix diverges at $Q=0$, i.e. $t^{-1}(0)=0$, or in explicitly form:
\be\label{eq:chi0}
1+g\chi(0)=0,\ T\le T_c.
\ee
It can be checked that equation Eq.(\ref{eq:chi0}) is exactly the same as the gap equation Eq.(\ref{eq:gap}).

In the same time, the mean-field self-energy can also be expressed in terms the T-matrix.
According to the Dyson equation, the self-energy is defined as the difference of $G_0(K)$ and $G(K)$ as follows
\be
\Sigma_\mf(K)&=&G_0^{-1}(K)-G^{-1}(K) \nn\\
&=&-\Delta_\mf^2i\sigma_y \tilde{G}_0(K)i\sigma_y.
\ee
In the superfluid phase, the T-matrix diverges at $Q=0$, which allow us to approximate it by a delta function located at zero momentum.
Therefore, we introduce the T-matrix for the condensate pairs
\be\label{eq:mf-t}
t_\mf(Q)=-\Delta_\mf^2\delta(Q),
\ee
Here the delta function of 4-momentum is $\delta(Q)=(1/T)\delta_{n,0}\delta^{(3)}(\bq)$,
then the self-energy can be rewritten as:
\be\label{eq:self-e}
\Sigma_\mf(K)&=&\sum_Q t_\mf(Q)i\sigma_y \tilde{G}_0(K-Q)i\sigma_y.
\ee
It seems that it is rather cumbersome to write the self-energy in such a form, but one can see that in this form, the full Green's function  can be treated as the bare propagator dressed by the condensate T-matrix.

We can see that only the contribution from the condensed pairs is considered in the BCS theory. In the BCS-BEC crossover, the attractive interaction gets much stronger when one approaches unitary limit or the BEC side, therefore one must also include the contribution from non-condensed pairs or pairing fluctuations. In contrast to the condensed pairs whose total momentum is zero, the non-condensed pairs have non-zero center of mass momentum and its effects become important if we want to consider the finite temperature physics. In the following section, we will consider the $G_0G$ pairing fluctuation theory, which allow us to generalize the above mean field theory results to the finite temperature with certain strong correlation effects included.

\section{pair fluctuation formalism}
\label{pair}

In the mean-field theory, only the contribution from condensed pairs was taken into account. We also want to include the contribution of non-condensed pairs at finite temperature. In general, the non-condensed pairs can be described by the amputated pair propagator or T-matrix with non-zero total momentum. Here we use the simple ladder diagrams to approximate the T-matrix. In momentum space, the T-matrix of non-condensed pairs are given by
\be
t_\pg(Q)=\frac{g}{1+g\chi(Q)},\ Q\not=0 \label{eq:t}
\ee
In order to be consistent with the BCS gap equation , we define the pair susceptibility as
\be\label{eq:chi}
\chi(Q)=\frac{1}{2\beta V}\mathrm{Tr}\sum_K \tilde{G}_0(K-Q)i\sigma_y G(K)i\sigma_y
\ee
as already discussed in \cite{KM,Ourreview}. The above T-matrix may looks exactly the same as the mean field case discussed in the section \ref{mf}, but here the full Green's function $G$ also contains the self-energy dressed by the non-condensed pairs which is much more complicated than the BCS full Green's function used in the section \ref{mf}.

The total T-matrix combining the condensed and non-condensed contribution is given by
\be
t(Q)=t_\mf(Q)+t_\pg(Q)
\ee
which give rise the following self energy
\be\label{eq:sig}
\Sigma(K)&=&\sum_Q t(Q)i\sigma_y \tilde{G}_0(K-Q)i\sigma_y\nn\\
&=&\Sigma_\mf(K)+\Sigma_\pg(K),
\ee
where
\be
&&\Sigma_\mf(K)=-\Delta_\mf^2 i\sigma_y \tilde{G}_0(K)i\sigma_y,\\
&&\Sigma_\pg(K)=\sum_Q t_\pg(Q)i\sigma_y \tilde{G}_0(K-Q)i\sigma_y,
\ee
In fact, the Eq.(\ref{eq:t}) and (\ref{eq:sig}) define a set of self-consistent equation system for the the unknown function $\Sigma_\pg$. With some initial guess of $\Sigma_\pg$, one can determine the full Green's function and further compute $t_\pg$. The T-matrix $t_{pg}$ can be used to determine $\Sigma_\pg$ again, which closed the loop. One can keep the above iteration until a convergence is reached. Once $\Sigma_\pg$ is obtained, the gap equation can be derived from generalized Thouless criterion
\be
t^{-1}(0)=\frac{1}{g}+\chi(0)=0\label{gap1}
\ee
However, it too difficult to solve the full $G_0G$ theory numerically.

In order to simplify the above mentioned numerical procedure, we have to make some further approximations.
Note from the Thouless criterion that $t_\pg$ is highly peaked around $Q=0$, hence the self-energy can be approximated as:
\be
\Sigma_\pg\approx-\Delta_\pg^2 i\sigma_y \tilde{G}_0(K)i\sigma_y,\ T\le T_c,
\ee
where the pseudogap is defined as:
\be\label{delpg}
\Delta_\pg^2=-\sum_{Q\neq0}t_\pg(Q).
\ee
After this approximation, the $\Sigma_\pg$ takes the same form as the BCS self-energy, which greatly simplifies calculations.
Then the total self-energy can also be written in the same form as BCS self-energy,
\be\label{self-energy}
\Sigma(K)=-\Delta^2 i\sigma_y \tilde{G}_0(K)i\sigma_y,
\ee
with $\Delta^2=\Delta_\mf^2+\Delta_\pg^2$. This also agrees with the intuitive picture that the correction of single particle propagator comes from both condensed and non-condensed pairs.

Since the total self-energy takes the same form as the BCS thoery, we expect that the gap equation, number equation, number difference equation also take the same form as in the BCS mean field theory, except that the order parameter $\Delta_\mf$ is replaced by total energy gap $\Delta$:
\be\label{gap}
\frac{\Delta}{g}&=&-\frac{1}{4}\sum_{\bk,\alpha}\frac{\partial E_\bk^\alpha}{\partial\Delta}[1-2f(E_\bk^\alpha)],\\
\label{num}
n&=&\frac{1}{2}\sum_{\bk,\alpha}\Big[1+\frac{\partial E_\bk^\alpha}{\partial\mu}[1-2f(E_\bk^\alpha)]\Big],\\
\label{dnum}
\delta n&=&\frac{1}{2}\sum_{\bk,\alpha}\frac{\partial E_\bk^\alpha}{\partial\dm}[1-2f(E_\bk^\alpha)].
\ee

To find out the pseudogap, we can compute the pair susceptibility by substituting the full Green's function $G(K)$ and free Green's function $G_0(K)$ into Eq.(\ref{eq:chi}). Note that $G(K)$ also take the same form as in BCS theory but with $\Delta_\mf$ replaced by $\Delta$. The result is
\begin{widetext}
\be
\chi(Q)&=&\frac{1}{2}\sum_{s,\gamma,\alpha=\pm}\sum_\bk\Big[\frac{s\gamma(\alpha\xi_\bk^2+\alpha\gamma\xi_\bk E_\bk^\alpha+\rho_\bk)[\lambda^2(k_x^2+k_y^2)-\lambda^2(k_x q_x+k_y q_y)]}{4\rho_\bk E_\bk^\alpha\eta_{\bk-\bq}}\nn\\
&&-\frac{\eta_{\bk-\bq}(\rho_\bk E_\bk^\alpha+\gamma\rho_\bk\xi_\bk+\alpha\gamma\xi_\bk\eta_\bk^2)+s\dm^2[\gamma\rho_\bk+\alpha\gamma(\Delta^2+\xi_\bk^2)+\alpha\xi_\bk E_\bk^\alpha]}{4\rho_\bk E_\bk^\alpha\eta_{\bk-\bq}}\Big]\nn\\
&&\times\frac{f(\xi_{\bk-\bq}^s)-f(\gamma E_\bk^\alpha)}{q_0-\gamma E_\bk^\alpha+\xi_{\bk-\bq}^s},
\ee
\end{widetext}
where $\xi_{\bk-\bq}^s=-\xi_{\bk-\bq}+s\eta_{\bk-\bq}$.

Then the pseudogap can be determined by Eq.(\ref{delpg}). In order to further simplify the calculations, we can take the advantage that the T-matrix is highly peaked around $Q=0$. Thus we expand the inverse T-matrix $t_{pg}^{-1}(Q)$ around $Q=0$ and only keep the leading terms. The gap equation Eq.(\ref{gap1}) ensures that constant term of this expansion is zero in the superfluid phase. Therefore, we find the following expansion
\be\label{eq:taylor}
t_{pg}^{-1}(Q)&\approx& a_0(i\Omega_n-\Omega_\bq)\\
\Omega_\bq&=&\sum_{i=1}^3B_iq_i^2\nn
\ee
Here we introduce the pair dispersion $\Omega_\bq$. $a_0$ is the inverse of spectral weight of fermion pairs and $B_i$ is the inverse of the effective mass of fermion pairs along direction $i$ with $i=x,y,z$.
We have:
\be
&&a_0=\frac{\partial \chi(Q)}{\partial q_0}\Big|_{Q=0}\nn\\
&&a_0B_i=-\frac{1}{2}\frac{\partial^2\chi(Q)}{\partial q_i^2}\Big|_{Q=0}\nn
\ee
where $i=x,y,z$. The detailed expressions of $a_0$ and $B_i$ are very complicated, which are written out explicitly in Appendix \ref{a0B}.

Inserting the above expansion to Eq.(\ref{delpg}), we find
\be\label{delpg1}
\Delta_{pg}^2&=&\frac{1}{a_0}\sum_\bq b(\Omega_\bq)
\ee
Here $b(x)=1/[\exp(x/T)-1]$ is the Bose distribution function. Solving Eq.(\ref{gap})-(\ref{dnum}) will determine the gap $\Delta$, chemical potential $\mu$, chemical potential difference $\delta\mu$ below $T_c$. With these results, the pseudogap $\Delta_\pg$ can be determined by Eq.(\ref{delpg1}) and the superfluid order parameter is given by $\Delta_\mf^2=\Delta^2-\Delta_\pg^2$. The vanishing of $\Delta_{sc}$ determines the critical temperature $T_c$, which is also equivalent to the condition $\Delta=\Delta_\pg$ at $T_c$.

Above $T_c$, there is no symmetry breaking, thus gap equation Eq.(\ref{gap1}) is not valid any more. We can extend our theory from below $T_c$ to above $T_c$ by insisting that in the normal phase the energy gap is solely generated by the pairing fluctuations. Thus we assume that $\Delta=\Delta_\pg$ for $T>T_c$. Since Eq.(\ref{gap1}) does not hold any more, the expansion of $t^{-1}_{pg}$ around $Q=0$ will give a non-zero constant term, which can be treat as the effective pair chemical potential.
\be
t_{pg}^{-1}(Q)\approx a_0\Big[i\Omega_n-(\Omega_\bq-\mu_b)\Big]
\ee
The pair chemical potential $\mu_b$ is determined by
\be
a_0\mu_b&=&\frac{1}{g}+\chi(0)\nn\\
&=&\frac{1}{g}+\frac{1}{4\Delta}\sum_{\bk,\alpha}\frac{\partial E_\bk^\alpha}{\partial\Delta}[1-2f(E_\bk^\alpha)]
\label{mub}
\ee
This equation will serve as the gap equation above $T_c$
Due to non-zero $\mu_b$, the pseudogap equation above $T_c$ is slightly modified as
\be\label{delpg2}
\Delta_{pg}^2&=&\frac{1}{a_0}\sum_\bq b(\Omega_\bq-\mu_b)
\ee
The number density and number density difference equation Eq.(\ref{num}),(\ref{dnum}) are the same as before. Combining the number equations with the new gap and pseudogap equations Eq.(\ref{mub}) and Eq.(\ref{delpg2}), one can determine $\mu,\ \dm,\ \mu_b,$ and $\Delta_\pg$ above $T_c$. Therefore we have generalized the pair fluctuation theory to the polarized Fermi gases with SOC. In the next section, we will show the numerical results of the above theory and also discuss its physical meaning.

\begin{figure*} \centering
\includegraphics[width=0.9\textwidth]{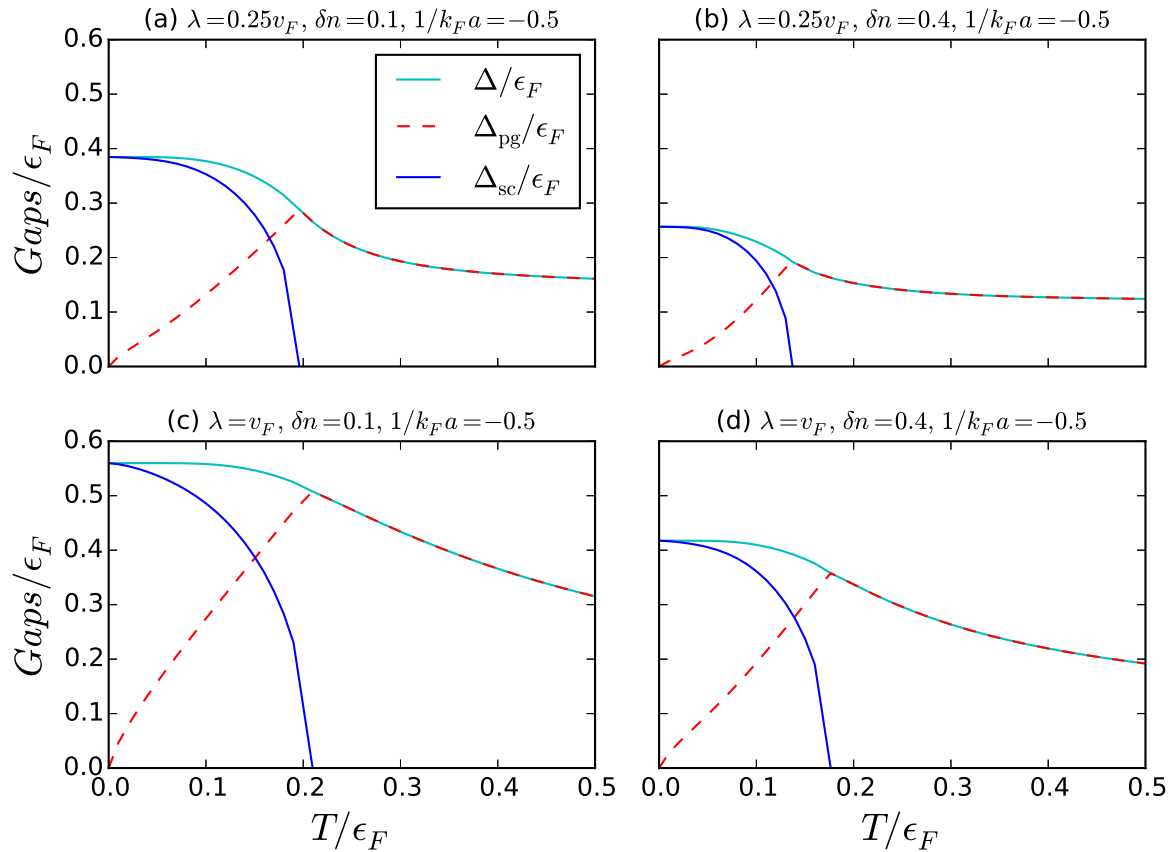}
\caption{The total gap $\Delta$, superfluid gap $\Delta_\mf$ and pseudogap $\Delta_\pg$ as functions of $T/E_F$ with interaction $1/k_Fa=-0.5$ which corresponds to the shallow BCS case. In panel (a),(b) and (c),(d), the SOC coupling constant $\lambda/v_F=0.25,\,1$ respectively. In panel (a),(c) and (d),(d), the number density difference $\delta n/n=0.1,\,0.4$ respectively.}
\label{fig:gap}
\end{figure*}

\begin{figure} \centering
\includegraphics[width=0.8\columnwidth]{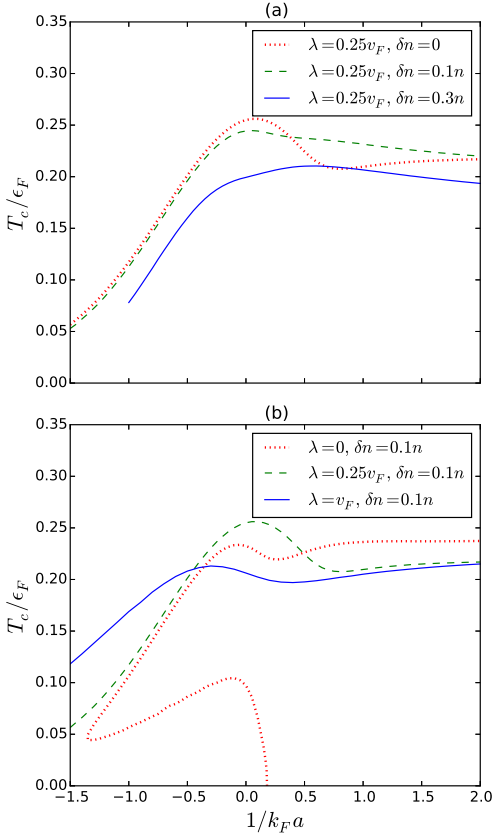}
\caption{(a) $T_c$ as a function of $1/k_Fa$, with fixed SOC couling $\lambda=0.25v_F$ and three different number density difference $\delta n/n=0,\,0.1,\,0.3$. (b) $T_c$ as a function of $1/k_Fa$, with fixed number density difference $\delta n/n=0.1$, with and three different SOC coupling $\lambda/v_F=0,\,0.25,\,1$.}
\label{fig:Tc}
\end{figure}

\begin{figure} \centering
\includegraphics[width=0.8\columnwidth]{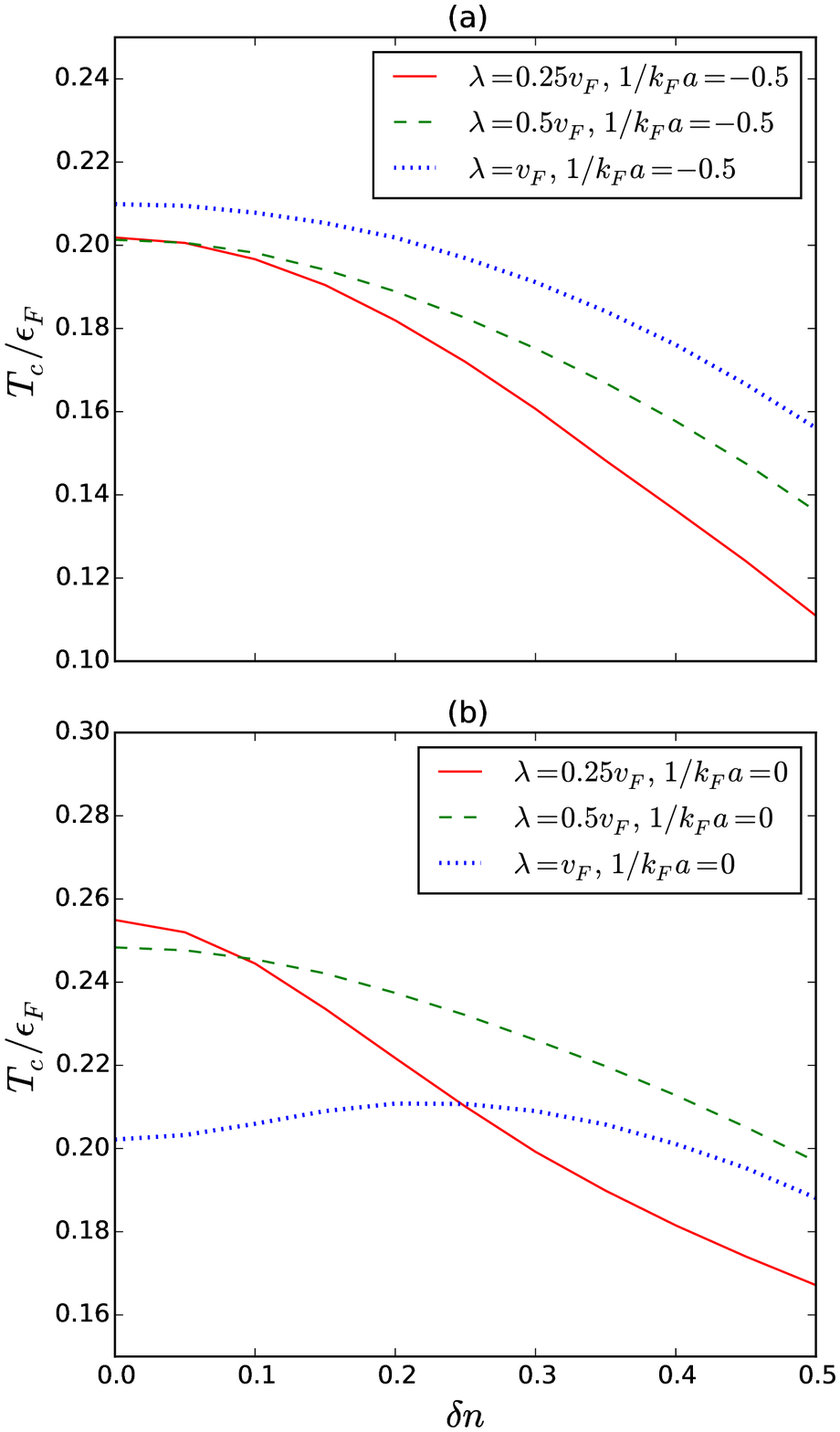}
\caption{(a) $T_c$ as a function of $\delta n/n$, with fixed $1/k_Fa=-0.5$ and three different SOC coupling $\lambda=0.25,\,0.5,\,1\,v_F$. (b) $T_c$ as a function of $\delta n/n$, with fixed $1/k_Fa=0$ and three different SOC coupling $\lambda/v_F=0.25,\,0.5,\,1$.}
\label{fig:Tc_dn}
\end{figure}

\section{numerical results and discussions}
\label{numerics}

In this section, we present our numerical results of the polarized Fermi gases with Rashba SOC. We will mainly focus on the polarized superfluid phase and the polarized pseudogap phase, and ignore the possibility of more exotic LOFF phase and phase separations. In the numerical calculations, the bare interaction strength $g$ is replaced by the experimental measurable scattering length $a$. These two quantities are related by the following regularization equation.
\be
\frac{1}{g}=\frac{m}{4\pi a}-\sum_\bk\frac{1}{2\epsilon_\bk}.
\ee
For convenience, we take the Fermi energy and momentum to be unity, which corresponds to taking the density as $n=1/3\pi^2$. We treat $1/(k_Fa)$ as the effective interaction.

Due to the strong attractive interaction, the appearance of pseudogap is quite general phenomena in the BCS-BEC crossover of Fermi gases. The advantage of our $G_0G$ pair fluctuation theory is that the pseudogap effect is relatively easy to compute. In Fig.\ref{fig:gap}, we first show the behaviors of total energy gap $\Delta$, superfluid gap $\Delta_\mf$ and pseudogap $\Delta_\pg$ depending on the temperature. All the plots in Fig. \ref{fig:gap} have similar features as the Fermi gases without the SOC and spin polarization. The total gap decrease with the increasing $T$ as in the mean field theory. $\Delta_{pg}$ is zero at $T=0$. Then, the pseudogap increases with increasing $T$ roughly as a power law function, reflecting the increasing of thermally excited non-condensed pairs.  Finally, the pseudogap saturates the whole energy gap, which determines the transition temperature $T_c$. Above $T_c$, the appearance of nonzero $\mu_b$ makes the pair number density become smaller, therefore the pseudogap gradually dies off. The polarized pseudogap phase is a special normal phase with single particle energy gap, which is clearly in contrast to the normal Fermi liquid. However, we should mention that above $T_c$, the pseudogap decays quite slow and will extend to even above $E_F$. But one should expect, there exist a crossover temperature $T^*$ above which the system should go back to Fermi liquid or Fermi gas. Therefore, the high $T$ part of the above plots only has qualitative meanings.

In the unitary limit and BEC side, the pairing effects are very strong. Thus the suppression of the superfluidity by the population imbalance is not very prominent, we can expect that in these regime the polarized superfluid phase is stable. On the other hand, in the deep BCS limit, one should expect that the polarized superfluid phase will be replaced by phase separation between balanced superfluid phase and normal phase of majority species. In Fig.\ref{fig:gap}, we shows the results for the shallow BCS case $1/(k_Fa)=-0.5$, which is roughly on the boundary between the stable and unstable polarized superfluid phase. Without the SOC, the pseudogap will become non-monotonic as a function of $T$ for the same set of parameters of $1/k_Fa$ and $\delta n/n$. In that case, there exists an upper and lower $T_c$, which clear signals that the polarized superfluid state are un-physical and there must be a mixture of superfluid state and normal state. In Fig.\ref{fig:gap}, we see that the introducing of SOC term increases the pairing effects and make the polarized superfluid state stable. Even when $\delta n/n=0.4$, the SOC coupling $\lambda=0.25v_F$ is still strong enough to stabilise the superfluid state. Comparing panel (c) and (d) to panel (a) and (b), we see that the increasing of SOC strength generally increase the overall magnitude of pairing gaps.

\begin{figure} \centering
\includegraphics[width=0.8\columnwidth]{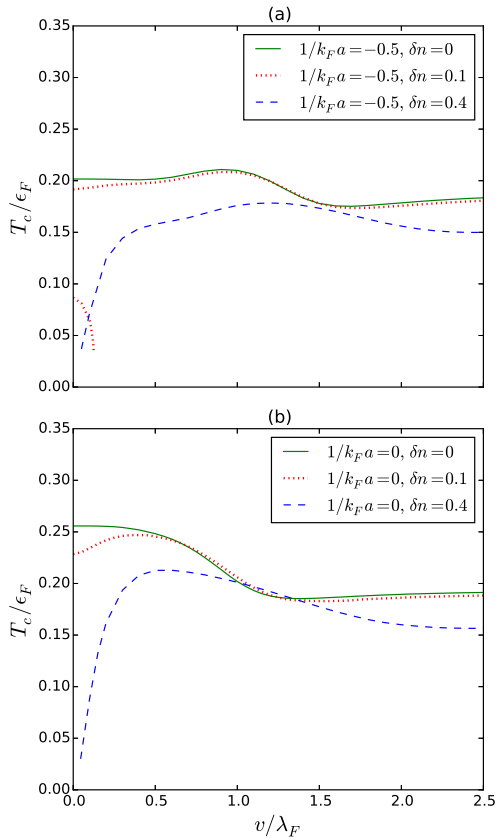}
\caption{(a) $T_c$ as a function of $\lambda/v_F$, with fixed $1/k_Fa=-0.5$ and three different number density difference $\delta n/n=0,\,0.1,\,0.4$. (b) Similar as panel (a) but with interaction $1/k_Fa=0$.}
\label{fig:Tc_lam}
\end{figure}

In Fig.\ref{fig:Tc}, we display the $T_c$ as a function of the interaction $1/k_Fa$, with different number density difference in panel (a) and with different SOC strength $\lambda$ in panel (b). For the balanced case with SOC, we reproduced the same result as in \cite{hu}. Comparing with the balanced Fermi gases without SOC, one can see that SOC term did not make too much qualitative change to the $T_c$ curve. The maximum of $T_c$ is reached around unitary limit, and there is a small local minimum located at the near BEC side. These features are quite general in the $G_0G$ pair fluctuation theory. If we consider the case with a small number density difference $\delta n/n=0.1$, we find that on the BCS side the $T_c$ is just slightly lower than that balanced case. But in the BEC side, this small number difference makes the small local minimum in the near BEC side disappear. Therefore, we are left with a little more broad maximum peak around the unitary limit. In the near BEC regime, the imbalanced Fermi gases have higher $T_c$ than the balanced case, in contrast to the expectation that the spin polarization should suppress superfluidity. This phenomenon actually already happened for very small $\delta n$ in the Fermi gases without SOC. The effect of SOC is to make this non-monotonic effects appear in a much larger parameter space than before. If we tune the number density difference to $\delta n/n=0.3$, then the whole $T_c$ curve is evidently lower than the balanced case. We also note that the location of the maximum is shifted to the BEC side which is same as the Fermi gases without SOC. For the $\delta n/n=0.3$ case, our $T_c$ curve stops at shallow BCS region because the un-physical behavior appears in a larger region due to the larger number difference.
In panel (b) of Fig.\ref{fig:Tc}, we show the $T_c$ curve with fixed number density difference $\delta n/n=0.1$. In the case without SOC, there appear the un-physical reentrant effects around the unitary and the BCS side as in \cite{Chien06}. However, when we turn on the SOC term, this un-physical behavior disappears. One can see that the SOC greatly stabilizes the polarized superfluid phase.

In Fig.\ref{fig:Tc_dn}, we show the $T_c$s as a function of number density difference $\delta n/n$, for different SOC couplings $\lambda/v_F=0.25,\,0.5,\,1$. The results of shallow BCS with $1/k_Fa=-0.5$ and unitary limit are presented in panel (a) and (b) respectively. In both cases, the increasing of $\delta n$ will rapidly suppress $T_c$ as one expected. As we discussed before, the effects of SOC is to enhance the pairing between fermions thus make the system more BEC like. Here we only present the results with large enough SOC such that there is no un-physical behavior in the $T_c$ curve. In the shallow BCS case, one can see that the increasing of $\lambda$ indeed push the whole $T_c$ curve upwards. On the other hand, we note that in the unitary limit, the SOC term actually make $T_c$ lower for small imbalance and increase $T_c$ for large imbalance. The overall effect of SOC in unitary is making the $T_c$ curve more flat than before. Similar behaviors are also observed in the BEC side. Therefore, for the unitary limit and the BEC side, the enhancement of pairing due to SOC roughly cancels out the suppression of pairing due to the population imbalance.

In Fig.\ref{fig:Tc_lam}, the $T_c$ as a function of $\lambda/v_F$ is plotted, with different number density imbalance $\delta n/n=0,\,0.1,\,0.4$, respectively. The results of shallow BCS with $1/k_Fa=-0.5$ and unitary limit are presented in panel (a) and (b) respectively. Since the SOC increases the pairing strength, the $T_c$ v.s. $\lambda/v_F$ plot can also be treated as another type of BCS-BEC crossover, with the large SOC corresponding to the deep BEC limit. When $\delta n=0$, we reproduce the results in \cite{hu}.
One can see that $T_c$ almost did not increase with the increasing of the SOC coupling. This is because we choose to study with shallow BCS or unitary limit in order to avoid un-physical behavior due to the imbalance. Close to the unitary limit, $T_c$ is already very high, thus obscure the effects of SOC. One can see that the $T_c$ v.s. $\lambda/v_F$ curve generally assume the similar shape as the $T_c$ v.s. $1/k_Fa$ plots. The $T_c$ reaches the highest value near the unitary limit then there is a small dip as one further increase the pairing strength. We should mention that there are some un-physical re-entrant behaviors appear in the small SOC area with large imbalance. Other than these small area, other part of the plot represents the stable polarized superfluid phase.

From the numerical results showing above, we can conclude that the SOC weaken the effect of the population imbalance.

\section{conclusion}
\label{conclu}

In this paper, we have studied the BCS-BEC crossover of three dimensional spin polarized Fermi gases with Rashba SOC. To capture the fluctuations due to the non-condensed pairs at finite temperature, we introduce a single particle self-energy dressed by the T-matrix or pair propagator approximated by $G_0G$ ladder diagrams. Based
on this $G_0G$ pair fluctuation theory, it is relatively straightforward to determine the pseudogap due to the stronger than BCS pairing. In the same time, the vanishing of the order parameter determines the $T_c$, which is beyond the simple mean field approximation.

With both SOC and spin polarization, there is a very large possible parameter space which may support a variety of exotic phases. In this paper, we focus on the simplest possible phase, the Sarma state or the spin polarized superfluid state, which is a direct generalization of BCS superfluid state. First, we have shown that the behaviors of the total gap and pseudogap with SOC are very similar to that without spin polarization, which suggests that the SOC increases the pairing effects and stabilize the spin polarized superfluid state.

We then present the detailed results of various $T_c$ curves which can also be thought as preliminary phase diagrams. One can see that without SOC, the un-physical re-entrant behavior appears in the shallow BCS side, which signals the phase separation between BCS superfluid state and spin polarized normal state. With certain amount of SOC coupling, one can check that spin polarized superfluid state are stabilized against the phase separation. From the $T_c$ v.s. $\delta n$ curve, one can see that the $T_c$ curves are flattened by increasing the SOC coupling, which implies that the effects of spin polarization and SOC largely canceled out. In summary, our work suggests that the stable region of spin polarized superfluid state are greatly enlarged by the introducing the Rashba SOC term.

The authors are supported by NSFC under grant No. 11404228.

\appendix
\section{detailed expressions for constant $a_0$ and $B_i$}
\label{a0B}

In the appendix, we show the detailed expressions for the spectral weight factor $a_0$ and inverse pair masses $B_i$ as follows.
\begin{widetext}
\be
a_0&=&\frac{\partial \chi(Q)}{\partial q_0}\Big|_{Q=0}\nn\\
&=&\frac{1}{2V}\sum_{s,\gamma,\alpha=\pm}\sum_\bk\Big[\frac{s\gamma(\alpha\xi_\bk^2+\alpha\gamma\xi_\bk E_\bk^\alpha+\rho_\bk)[\lambda^2(k_x^2+k_y^2)]}{4\rho_\bk E_\bk^\alpha\eta_{\bk}}\nn\\
&&-\frac{\eta_{\bk}(\rho_\bk E_\bk^\alpha+\gamma\rho_\bk\xi_\bk+\alpha\gamma\xi_\bk\eta_\bk^2)+s\dm^2[\gamma\rho_\bk+\alpha\gamma(\Delta^2+\xi_\bk^2)+\alpha\xi_\bk E_\bk^\alpha]}{4\rho_\bk E_\bk^\alpha\eta_{\bk}}\Big]\nn\\
&&\times\frac{-f(\xi_{\bk}^s)+f(\gamma E_\bk^\alpha)}{(\xi_{\bk}^s-\gamma E_\bk^\alpha)^2},
\ee
\be
a_0B_z&=&-\frac{1}{2}\frac{\partial^2\chi(Q)}{\partial q_z^2}\Big|_{Q=0}\nn\\
&=&-\frac{1}{4V}\sum_{s,\gamma,\alpha=\pm}\sum_\bk\Big[\frac{s\gamma(\alpha\xi_\bk^2+\alpha\gamma\xi_\bk E_\bk^\alpha+\rho_\bk)[\lambda^2(k_x^2+k_y^2)]}{4\rho_\bk E_\bk^\alpha\eta_{\bk}}\nn\\
&&-\frac{\eta_{\bk}(\rho_\bk E_\bk^\alpha+\gamma\rho_\bk\xi_\bk+\alpha\gamma\xi_\bk\eta_\bk^2)+s\dm^2[\gamma\rho_\bk+\alpha\gamma(\Delta^2+\xi_\bk^2)+\alpha\xi_\bk E_\bk^\alpha]}{4\rho_\bk E_\bk^\alpha\eta_{\bk}}\Big]\nn\\
&&\times\Big[\frac{2\beta k_z^2f(\xi_\bk^s)[1-f(\xi_\bk^s)]}{m^2(\xi_\bk^s-\gamma E_\bk^\alpha)^2}+\frac{f(\xi_\bk^s)[1-f(\xi_\bk^s)][m\beta+k_z^2\beta^2(1-2f(\xi_\bk^s))]}{m^2(\xi_\bk^s-\gamma E_\bk^\alpha)}\nn\\
&&+[f(\xi_\bk^s)-f(\gamma E_\bk^\alpha)]\big[\frac{2k_z^2}{m^2(\xi_\bk^s-\gamma E_\bk^\alpha)^3}+\frac{1}{m(\xi_\bk^s-\gamma E_\bk^\alpha)^2}\big]\Big],
\ee
\be
a_0B_i&=&-\frac{1}{2}\frac{\partial^2\chi(Q)}{\partial q_i^2}\Big|_{Q=0}\nn\\
&=&-\frac{1}{4V}\sum_{s,\gamma,\alpha=\pm}\sum_\bk\Big\{2\Big[\frac{\lambda^2k_i[f(\xi_\bk^s)-f(\gamma E_\bk^\alpha)]}{4E_\bk^\alpha \eta_\bk^3}-\frac{\beta f(\xi_\bk^s)(1-f(\xi_\bk^s))J}{4E_\bk^\alpha\eta_\bk}\Big]\nn\\
&&\times\Big[-\frac{JL}{\rho_\bk(\xi_\bk^s-\gamma E_\bk^\alpha)^2}+\frac{\lambda^2 k_i(C-A\eta_\bk)}{\rho_\bk\eta_\bk(\xi_\bk^s-\gamma E_\bk^\alpha)}\Big]\nn\\
&&+\frac{L}{4\rho_\bk E_\bk^\alpha(\xi_\bk^s-\gamma E_\bk^\alpha)}\Big[(f(\xi_\bk^s)-f(\gamma E_\bk^\alpha))\big(\frac{3\lambda^4 k_i^2}{\eta_\bk^5}-\frac{\lambda^2}{\eta_\bk^3}\big)-\frac{2\beta\lambda^2 k_i f(\xi_\bk^s)(1-f(\xi_\bk^s))J}{\eta_\bk^3}\nn\\
&&+\frac{1}{\eta_\bk}\big[-\beta f(\xi_\bk^s)(1-f(\xi_\bk^s))(-\frac{1}{m}-\frac{s\lambda^4 k_i^2}{\eta_\bk^3}+\frac{s\lambda^2}{\eta_\bk})+2\beta^2 f(\xi_\bk^s)(1-f(\xi_\bk^s))(1-2f(\xi_\bk^s)) J^2\big]\Big]\nn\\
&&+\frac{f(\xi_\bk^s)-f(\gamma E_\bk^\alpha)}{4\rho_\bk E_\bk^\alpha\eta_\bk}\Big[-\frac{2\lambda^2 k_i(C-A\eta_\bk)J}{\eta_\bk(\xi_\bk^s-\gamma E_\bk^\alpha)^2}-\frac{\lambda^2 C(\eta_\bk-\lambda^2 k_i^2)}{\eta_\bk^3(\xi_\bk^s-\gamma E_\bk^\alpha)}\nn\\
&&+L\big[\frac{2J^2}{(\xi_\bk^s-\gamma E_\bk^\alpha)^3}-\frac{1}{(\xi_\bk^s-\gamma E_\bk^\alpha)^2}\big(-\frac{1}{m}-\frac{s\lambda^4 k_i^2}{\eta_\bk^3}+\frac{s\lambda^2}{\eta_\bk}\big)\big]\Big]\Big\}.
\ee
where $i=x,y$ and
\be
&&A=s\gamma(\alpha\xi_\bk^2+\alpha\gamma\xi_\bk E_\bk^\alpha+\rho_\bk),\qquad B=\lambda^2(k_x^2+k_y^2)\nn\\
&&C=\rho_\bk E_\bk^\alpha+\gamma\rho_\bk\xi_\bk+\alpha\gamma\xi_\bk\eta_\bk^2,
\qquad D=s\dm^2[\gamma\rho_\bk+\alpha\gamma(\Delta^2+\xi_\bk^2)+\alpha\xi_\bk E_\bk^\alpha]\nn\\
&&J=\frac{k_i}{m}-\frac{s\lambda^2 k_i}{\eta_\bk},\qquad L=AB-D-C\eta_\bk\nn\\
&&\xi_{\bk-\bq}^s=-\xi_{\bk-\bq}+s\eta_{\bk-\bq}
\ee

\end{widetext}


\end{document}